\documentclass[eqsecnum,aps,prd,nofootinbib]{revtex4}

\usepackage{amsmath}
\usepackage{enumerate}
\usepackage{amsfonts}

\begin{document}

\title {Energy Conditions and Junction Conditions}
\author{Donald Marolf}

\affiliation{Physics Department, UCSB, Santa Barbara, CA 93106, USA\\
and \\
Perimeter Institute, 31 Caroline St. N, Waterloo,
Ontario N2L 2Y5, Canada \\
{\tt marolf@physics.ucsb.edu}}

\author{Sho Yaida}
\affiliation{Mathematics Department, UCSB, Santa Barbara, CA 93106, USA\\
{\tt yaida@umail.ucsb.edu}}

\date{August, 2005}

\begin{abstract}
We consider the familiar junction conditions described by Israel for
thin timelike walls in Einstein-Hilbert gravity.  One such condition
requires the induced metric to be continuous across the wall.  Now,
there are many spacetimes with sources confined to a thin wall for
which this condition is violated and the Israel formalism does not
apply. However, we explore the conjecture that the induced metric is
in fact continuous for any thin wall which models spacetimes
containing only positive energy matter.  Thus, the usual junction
conditions would hold for all positive energy spacetimes. This
conjecture is proven in various special cases, including the case of
static spacetimes with spherical or planar symmetry as well as
settings without symmetry which may be sufficiently well
approximated by smooth spacetimes with well-behaved null geodesic
congruences.
\end{abstract}

\maketitle

\section{Introduction}
\label{intro}

The study of boundary layers (i.e., singular sheets of sources with
zero thickness) is of longstanding interest in both electromagnetism
and general relativity.  In addition to approximating smooth
solutions such as domain walls, the class of ``thin-wall'' solutions
provides a useful laboratory in which to explore new phenomena.  For
example, thin-wall spacetimes have been of much use in investigating
so-called ``braneworld'' scenarios, first suggested in the modern
context in \cite{BW1,BW2,BW3}, in which the visible 3+1 universe is
confined to a submanifold of some higher dimensional spacetime. In
particular, the phenomenon by which gravity  can become localized
along such a domain wall was discovered by Randall and Sundrum
\cite{RS1,RS2} by considering the limit of an infinitely thin wall.
Thin-wall spacetimes are also historically of interest in exploring
gravitational collapse (see e.g. \cite{I2}), typically in the
context of spherical shells.

Although the stress tensor must diverge at an infinitely thin wall,
the associated singularities in Einstein-Hilbert gravity are often
mild. For familiar cases, these singularities serve merely to
simplify the equations of motion by turning the differential
equations for the fields into finite difference equations, known as
``junction conditions,'' governing the discontinuities of various
fields across the thin wall. The standard formalism for treating
thin walls with spacelike normals was developed by Israel
\cite{Israel}, building on the work of others (e.g.,
\cite{CL1,CL2,GD,MS}).

Briefly, the Israel formalism \cite{Israel} considers spacetimes
with a distinguished singular hypersurface $\Sigma^0$ of
co-dimension one.  One imagines foliating the spacetime near
$\Sigma^0$ in an arbitrary way such that $\Sigma^0$ is one of the
hypersurfaces in the foliation.  One then requires that both the
induced metric $h$ and the extrinsic curvature $K$ of
hypersurfaces in this foliation have well-defined limits $h_\pm,
K_\pm$ as the singular hypersurface $\Sigma^0$ is approached from
each side. Furthermore, one requires that the induced metric is
continuous across $\Sigma^0$: $h_+ = h_-.$  In such a setting,
Israel showed that the remaining junction condition for a wall
with spacelike normal is

\begin{equation}
\label{Kjunction} [K]_{ij} =  8 \pi G_N \left( S_{ij} -
\frac{h_{ij}}{d-2} h^{kl} S_{kl} \right).
\end{equation}
Here $[K]_{ij} = K_{+ij} - K_{-ij}$ is the discontinuity in the
extrinsic curvature across the surface  and $S_{ij}$ is the
so-called surface stress tensor, which is essentially the
pull-back of the stress-energy tensor integrated over a small
region around the hypersurface $\Sigma^0$.  We work in $d$
spacetime dimensions (so that the wall has $(d-1)$ spacetime
dimensions) and $G_N$ is Newton's  gravitational constant. In this
formalism, components of the stress tensor not captured by
$S_{ij}$ remain bounded at the wall, so that the tensor $S_{ij}$
captures the ``delta-function part'' of the stress-energy tensor.

Since condition (\ref{Kjunction}) follows from the Einstein equation
integrated over a small region,  it is firmly grounded in physics.
Consider, however, the more fundamental requirement of continuity of
the induced metric: $h_+ = h_-$.  In textbooks, this requirement is
often motivated by aesthetic concerns, such as the desire for
$\Sigma^0$ to have a well-defined induced geometry, or the related
desire to have a simple description of the dynamics of particles
bound to $\Sigma^0$. Another motivation is that the ansatz $h_+=h_-$
does in fact allow one to solve the Einstein equation (via
(\ref{Kjunction})) whenever the singular part of the stress tensor
is captured by $S_{ij}$. What remains unclear, however, is whether
this requirement is in fact satisfied by all cases of physical
interest.

Israel's own motivations  \cite{Ichat} for imposing the continuity
of $h$ are somewhat more enlightening, and were based on an
analogy with electromagnetism (in particular, with
electrostatics).  In the electrostatic context, one finds that a
discontinuity in the component $E_\perp$  of the electric field
orthogonal to the surface is associated with a surface charge
density $\sigma$ in direct correspondence with (\ref{Kjunction}).
In contrast, a discontinuity in the parallel components
$E_\parallel$ of the electric field is associated with a
distribution of {\it dipole} sources on the surface.  As one
expects no gravitational point dipoles, one might also expect the
gravitational field (and in particular the induced metric) to be
continuous across the surface. Of course, the reason that one
expects no gravitational dipoles is that point dipoles require
both positive and negative charges.

Our purpose in the present work is to explore the conjecture
(implicit in the above reasoning) that taking thin-wall limits of
spacetimes satisfying an appropriate positive energy condition will
necessarily lead to continuity of the induced metric $h$, and thus
to the familiar condition (\ref{Kjunction}). Such a conjecture is
non-trivial due to the intrinsic non-linearities of general
relativity.  Indeed, since the curvature diverges at thin walls, one
expects non-linearities to have significant effects.

Let us take a few moments to make the notion of a ``thin-wall
limit'' more precise. To do so, consider a one-parameter family of
smooth Lorentz-signature metrics $g_\lambda$ on a fixed manifold $M$
for $\lambda \in (0,1]$. Thus, each $(M,g_\lambda)$ is a spacetime.
However, we think of each $g_\lambda$ as merely some tensor field on
the manifold $M$, where in particular the differential structure of
$M$ is fixed and independent of $\lambda$. We require the spacetime
dimension of $M$ to be $d \ge 3$ so that the Einstein equation
contains non-trivial dynamics.

In the limit $\lambda \rightarrow 0$, we wish to allow a thin wall
to develop on some hypersurface; that is, on a smooth embedded
submanifold of co-dimension one, say $\Sigma^0$. Note that, while we
require $\Sigma^0$ to be smooth with respect to the differentiable
structure of $M$, we will {\it not} require the limit $\lim_{\lambda
\rightarrow 0} g_\lambda$ of the tensor fields $g_\lambda$ to be
smooth there. Thus, we allow a thin wall to form at $\Sigma^0$ as
one decreases the width of any  ``thick wall'' which may be present
for $\lambda \neq 0$.

However, we will require the limiting tensor field to be smooth away
from $\Sigma^0$ so that we encounter only thin-wall singularities.
In particular, we suppose that our family $g_\lambda$ satisfies the
following conditions:

\begin{enumerate}[i)]
\item \label{smooth} The limit of $g_\lambda$ as $\lambda
\rightarrow 0$ defines a smooth Lorentz-signature metric $g_0$
everywhere away from $\Sigma^0$.  Similarly, we require that the
first and second derivatives of $g_\lambda$ (with respect to any
fixed smooth coordinate) converge to the first and second
derivatives of $g_0$ everywhere on $M$ away from $\Sigma^0$.  For
simplicity, we have in mind taking the pointwise limit of the
components of each tensor field $g_\lambda$ and of its derivatives.
Thus, this assumption restricts not only the geometry associated
with $g_\lambda$, but also the inherent freedom to perform
diffeomorphisms as a function of $\lambda$. Note that we do not
require $(M,g_\lambda)$ to be in any sense complete, so that $M$ may
in fact represent a small region of a larger manifold and other
singularities could arise elsewhere.

\item \label{ncc} Each $g_\lambda$ with $\lambda > 0$ satisfies
the null convergence condition: $R_{ab} k^a k^b \ge 0$ for all null
$k^a$.
 Here our conventions are those of \cite{Wald}.
\end{enumerate}

Note that, without condition (\ref{ncc}), it is straightforward to
construct a family $g_\lambda$ satisfying (\ref{smooth}) such that
the induced metric in the limit $\lambda \rightarrow 0$ is
discontinuous across the wall. For example, one may consider the
following metrics on $\mathbb{R}^d$:

\begin{equation}
ds^2_\lambda := (g_\lambda)_{ab} dx^a dx^b =
-f_\lambda(z)dt^2+dz^2+\mathop{\sum_{i=1}^{d-2}}{(dx^i)^2},
\end{equation}
where $f_{\lambda}$ is some smooth (positive) function with
$f_{\lambda}(z)=1$ for $z \le -\lambda$ and $f_{\lambda}(z)=2$ for
$z \ge +\lambda$. This one-parameter family of metrics satisfies the
condition (\ref{smooth}), but, as one may check, it violates the
null convergence condition (\ref{ncc}).

Our goal is to investigate what further technical assumptions must
be added to (\ref{smooth}) and (\ref{ncc}) in order to guarantee
that $g_0$ defines an induced metric on $\Sigma^0$ which is
continuous across the wall.  While a completely general theorem is
beyond the scope of this work, we present some partial results
below.  In section \ref{proof}, we prove a number of results to the
effect that, in the context of various ans\"atze for the metrics
$g_\lambda$, the conditions (\ref{smooth}) and (\ref{ncc}) do indeed
imply that $g_0$ defines an induced metric which is continuous
across $\Sigma^0$.  In particular, section \ref{plane} considers a
case where one assumes translation invariance along the surface
$\Sigma^0$ as well as rotation symmetry among the spatial directions
of $\Sigma^0$.    This case may also be viewed as the warped product
of a real line with another real line (the time direction) and with
a Euclidean plane, where warping of the second and third factors is
allowed only with respect to the first real line.  This simple
setting also serves to introduce a few technical tools (Lemmas)
which will continue to be of use later in section \ref{warp}, where
we allow the plane in the above warped product to be replaced with
an arbitrary geometry; i.e., we generalize to metrics of the form

\begin{equation}
\label{w1} ds^2 = -e^{2\alpha
(z)}dt^2+dz^2+e^{2\beta(z)}\mathop{\sum_{i, j=1}^{d-2}} \tilde
g_{ij}(x){dx^i}{dx^j},
\end{equation}
where $x = (x^1,...,x^{d-2})$ is an appropriate set of spatial
coordinates on the hypersurfaces of the foliation. In particular,
any given static, spherically symmetric metric may be written in the
form (\ref{w1}). For this case, we also introduce a further
assumption guaranteeing that the wall is in fact ``thin'' in a sense
measured by certain null geodesics. A final static special case with
more degrees of freedom (but of less physical interest) is studied
in section \ref{diagonal}.

In contrast, section \ref{congruence} pursues a different approach.
There we prove a theorem which requires neither symmetry nor a
particular ansatz for the metric. However, the price to be paid is
the introduction of a number of detailed assumptions which relate
the behavior of null geodesic congruences in the spacetime $(M,g_0)$
to that of null geodesic congruences in $(M,g_\lambda)$. We also
assume that the induced metric on each side of the thin wall is
well-defined, whereas we were able to derive this result within the
context of section \ref{proof}. Finally, section \ref{congruence}
requires the induced metric on each side to be timelike.  With these
additional assumptions, the Raychaudhuri equation leads directly to
continuity of the induced metric. Through its use of the
Raychaudhuri equation, this theorem gives some physical insight into
how the null convergence condition helps to ensure this continuity.
We then close with a brief discussion in section \ref{disc}. In
particular, we emphasize that if one does {\it not} impose a
positive energy condition (e.g., as in the Randall-Sundrum scenario
\cite{RS1,RS2}), then there is in general no reason to expect the
induced metric to be continuous across the junction.

\section{Proofs of the conjecture in special cases}

\label{proof}

We now turn to proofs of the above conjecture in various special
cases. We shall begin by considering (section \ref{plane}) a special
case of the setting laid out in section \ref{intro} above which has
a particularly high degree of symmetry. This will simplify the
algebra involved and allow us to introduce two Lemmas in a context
where their use is transparent. We then proceed in section
\ref{warp} to a more general static case involving a warped product,
as would be appropriate for spherical symmetry. In section
\ref{diagonal}, we briefly examine a further special case with more
degrees of freedom, but which is of less physical interest as it
requires a timelike Killing field of constant norm.

\subsection{Translation and Spatial rotation invariance: a simple first case}

\label{plane}

We begin with the case where the metrics $g_\lambda$ are static and
share a common set of Euclidean (i.e., rotational and translational)
symmetries. We further assume that Gaussian normal coordinates in
$(M,g_\lambda)$ based on the hypersurface $\Sigma^0$ are independent
of $\lambda$. Thus, our metrics take the form:
\begin{equation}
\label{simple} ds^2_\lambda =
-e^{2\alpha_{\lambda}(z)}dt^2+dz^2+e^{2\beta_{\lambda}(z)}\mathop{\sum_{i=1}^{d-2}}{(dx^i)}^2,
\end{equation}
where $x = (x^1,...,x^{d-2})$ provides an appropriate coordinate
system on the $(d-2)$-planes which form the orbits of the Euclidean
symmetry. This case includes, for example, the setting studied by
Randall and Sundrum \cite{RS1,RS2}.  For definiteness, we take the
range of $z$ to include the $\lambda$-independent closed interval
$[z_-,z_+]$ with $z_+ > 0 > z_-$, and we will always work on this
closed interval below.

Here, condition (\ref{smooth}) of section \ref{intro} merely imposes
that $\alpha_\lambda$, $\beta_\lambda$, and their first and second
derivatives converge for $z \neq 0$ to some $\alpha_0,\beta_0$ (and
their derivatives) in the limit $\lambda \rightarrow 0$. On the
other hand, requirement (\ref{ncc}) imposes the null convergence
condition for each $\lambda$. Considering the null vectors
\begin{eqnarray}
k_\perp &=&  e^{-\alpha_\lambda} \partial_t + \partial_z, \cr
k_\parallel &=&  e^{-\alpha_\lambda} \partial_t + e^{-\beta_\lambda}
\partial_{x_1},
\end{eqnarray}
one finds:
\begin{eqnarray}
\label{perp} 0 &\le& R_{ab} k^a_\perp k^b_\perp  =
(d-2)[-{\beta''_{\lambda}}-{\beta'_{\lambda}}^2+\alpha'_{\lambda}\beta'_{\lambda}],
\ \ \ {\rm and}
\\ \label{para} 0 &\le& R_{ab} k^a_\parallel k^b_\parallel =
{\alpha''_{\lambda}}+{\alpha'_{\lambda}}^2+(d-3)\alpha'_{\lambda}\beta'_{\lambda}-{\beta''_{\lambda}}-(d-2){\beta'_{\lambda}}^2.
\end{eqnarray}

We now wish to prove continuity of $\alpha_0,\beta_0$ at $z=0$.
The essential strategy will be to use (\ref{perp}), (\ref{para}) and
the convergence of $\alpha_\lambda, \beta_\lambda$ to  $\alpha_0,
\beta_0$ (for $z\neq 0$) to show that the derivatives
$\alpha'_\lambda, \beta'_\lambda$ at $z=0$ satisfy a bound that is
independent of $\lambda$ for small $\lambda$.  Thus, even the
limiting functions $\alpha_0, \beta_0$ must be continuous at $z=0$.

However, before commencing our main argument, we first note that
condition (\ref{perp}) restricts the possible local extrema of
$\beta_\lambda$. Indeed, since the first derivative vanishes at such
an extremum, we find $\beta_\lambda '' \leq 0$.  If
$\beta_\lambda''$ is non-vanishing there, the extremum must be a
local maximum.

In fact, the following argument shows that there are no local minima
even in the case where $\beta_\lambda''$ vanishes at an extremum.
Suppose $\beta_\lambda$ satisfies (\ref{perp}) and has a local
minimum at $z_{min}$.  Then $\beta_\lambda'$ is negative somewhere
before $z_{min}$, and we may in fact choose an open interval on
which $\beta_\lambda'$ is negative, but such that $\beta_\lambda'$
approaches zero at the right end of this interval\footnote{This
endpoint need not be $z_{min}$.}.    On that open interval we may
write (\ref{perp}) as

\begin{equation}
0\leq    \frac{ 1}{(-\beta_\lambda')} [-
\beta''_{\lambda}-(\beta'_{\lambda})^2+ \alpha'_{\lambda}
\beta'_{\lambda} ] = \frac{\beta''_{\lambda}}{\beta'_{\lambda}}-
(\alpha'_{\lambda}-\beta'_{\lambda}) =
\frac{d}{dz}(\log(-\beta'_{\lambda}))-
(\alpha'_{\lambda}-\beta'_{\lambda}).
\end{equation}
But $\beta'_\lambda$ vanishes at the endpoint, so $\log
(-\beta'_\lambda)$ diverges toward negative infinity.  On the other
hand, since $\alpha_\lambda$ and $\beta_\lambda$ are smooth
functions, the remaining term $\alpha'_{\lambda}-\beta'_{\lambda}$
remains bounded.  Thus, we have reached a contradiction and
$\beta_\lambda$ can have no local minima. A similar argument using
equation (\ref{para}) in the form
\begin{equation}
-\frac{d^2}{dz^2}(\beta_{\lambda}-\alpha_{\lambda})+(\beta'_{\lambda}-\alpha'_{\lambda})[-\alpha'_{\lambda}-(d-2)\beta'_{\lambda}]\geq0
\end{equation}
shows that the function $\beta_\lambda - \alpha_\lambda$ can have no
local minima.

Let us summarize these results and some of the implications in the
following Lemma:

\begin{quote}

$  Lemma \ 1$:  Suppose that, for some $\lambda$, equations
(\ref{perp}) and (\ref{para}) hold on an interval $[z_-,z_+]$ and
that $\alpha_\lambda$, $\beta_\lambda$ are smooth on this interval.
Then we have the lower bounds $\beta_\lambda \ge
min\{\beta_{\lambda}(z_{-}), \beta_{\lambda}(z_{+})\}$ and
$\beta_\lambda - \alpha_\lambda \ge min\{(\beta_{\lambda}-
\alpha_\lambda)(z_{-}), (\beta_{\lambda}- \alpha_\lambda)(z_{+})\}$
everywhere on this interval.
\end{quote}

Now, as a result of the assumption that the $\beta_\lambda$ converge
to $\beta_0$ for $z \neq 0$, the values $\beta_\lambda(z_\pm)$ are
close to $\beta_0 (z_\pm)$ for small $\lambda$ (and similarly for
$\beta_\lambda - \alpha_\lambda$). Thus we also have

\begin{quote}
$  Corollary \ 1$:  The functions ${\beta_{\lambda}}$ and
${\beta_{\lambda}-\alpha_{\lambda}}$ are bounded below on the
interval $[z_-,z_+]$, uniformly in $\lambda$ for small $\lambda$. As
a direct consequence, for small $\lambda$ the functions
$e^{-\beta_{\lambda}}$ and $e^{\alpha_{\lambda}- \beta_\lambda}$ are
bounded (uniformly in $\lambda$) both above and below.
\end{quote}

To simplify the remaining analysis, it is useful to introduce the
following two quantities:
\begin{equation}
\label{AB} {A_\lambda} :=-
\beta'_{\lambda}e^{\beta_{\lambda}-\alpha_{\lambda}}, \ \ \
{B_\lambda}
:=[\frac{d}{dz}(e^{\alpha_{\lambda}-\beta_{\lambda}})]e^{(d-1)\beta_{\lambda}}.
\end{equation}
With these definitions, (\ref{perp}) and (\ref{para}) are equivalent
to:
\begin{eqnarray}
0\leq\frac{d}{dz}[{A_\lambda} ] \ \ \ {\rm and} \ \ \
0\leq\frac{d}{dz}[{B_\lambda} ],
\end{eqnarray}
so that both ${A_\lambda} $ and ${B_\lambda} $ are non-decreasing.
In particular, ${A_\lambda} (z_{-})\leq {A_\lambda} (z)\leq
{A_\lambda} (z_{+})$ for  $z \in [z_-,z_+]$, and similarly for
${B_\lambda} $. Since, for small $\lambda$, ${A_\lambda} (z_\pm)$
is close to ${A_0} (z_\pm)$ and ${B_\lambda} (z_\pm)$ is close to
${B_0} (z_\pm)$, we arrive at the following Lemma:

\begin{quote}
$  Lemma \ 2$:  The functions ${A_\lambda} $ and ${B_\lambda} $ are
bounded both above and below on the interval $[z_-,z_+]$.
Furthermore, for sufficiently small $\lambda$, one may choose bounds
that are independent of $\lambda$.
\end{quote}

The proof of the conjecture now follows by combining Corollary  1
and Lemma 2.  We see  that both $\beta_\lambda' = -{A_\lambda}
e^{\alpha_\lambda - \beta_\lambda}$ and
$\frac{d}{dz}(e^{\alpha_{\lambda}-\beta_{\lambda}}) = {B_\lambda}
e^{-(d-1)\beta_{\lambda}}$ are bounded uniformly in $\lambda$ for
small $\lambda$, so that $\beta_0$, $e^{\alpha_{0}-\beta_{0}}$, and
therefore $e^{\alpha_0}$ must be continuous.   Let us summarize the
result in the following Theorem:

\begin{quote}
$  Theorem \ 1$: For a one-parameter family of smooth metrics
described by the ansatz (\ref{simple}) and satisfying conditions
(\ref{smooth}) and (\ref{ncc}) from section \ref{intro}, the limiting metric $g_0$ defines continuous induced metrics on the hypersurfaces $\Sigma^z$. \\
\end{quote}
Thus, we have proven our conjecture for the special case described
by the ansatz (\ref{simple}).

\subsection{More general Warped products}

\label{warp}

We now consider the more general ansatz:
\begin{eqnarray}
\label{warped} ds_\lambda^2 =
-e^{2\alpha_{\lambda}(z)}dt^2+dz^2+e^{2\beta_{\lambda}(z)}\mathop{\sum_{i,
j=1}^{d-2}}\tilde g_{ij}(x){dx^i}{dx^j},
\end{eqnarray}
where $\tilde g_{ij}$ is independent of $t$ and $z$, and where $x =
(x^1,...,x^{d-2})$ is an appropriate collection of additional
coordinates. Again, we take the range of $z$ to include some closed
interval $[z_-,z_+]$, and we confine all discussion to this interval
below. This ansatz includes static spherically symmetric families of
metrics $g_\lambda$ when $\tilde g_{ij}$ is the metric on the round
unit sphere $S^{d-2}$.

An interesting feature of such spacetimes is that geodesics which
are initially tangent to a surface $x^i = constant$ in fact remain
in this surface.  One may show that affinely parameterized null
geodesics of this sort are integral curves of the vector fields
$e^{-2\alpha_\lambda}
\partial_t \pm e^{-\alpha_\lambda} \partial_z$.

Thus each such null geodesic requires an affine parameter

\begin{equation}
\label{affine} \mathop{\int_{z_-}^{z_+}}e^{\alpha_\lambda}dz
\end{equation}
to traverse the region between $z_-$ and $z_+$. We see that if the
two sides of the wall are to remain in causal contact (i.e., if
information can flow across the wall in {\it either} direction) in
the limit $\lambda \rightarrow 0$, the above quantity must be
bounded uniformly in $\lambda$ for small $\lambda$. We therefore
assume such a uniform bound in our treatment below.

Let us now proceed with the proof.  First, we note that the
consequences of condition (\ref{smooth}) are identical to those in
section \ref{plane}; namely, that away from $z=0$ the functions
$\alpha_\lambda$, $\beta_\lambda$, and their first and second
derivatives converge to some $\alpha_0,\beta_0$ (and their
derivatives) in the limit $\lambda \rightarrow 0$. Let us consider
the null convergence condition (requirement (\ref{ncc})) for the
null vectors
\begin{eqnarray}
k_\perp &=&  e^{-\alpha_\lambda} \partial_t + \partial_z, \ \ \ {\rm
and} \cr k_\parallel &=&  e^{-\alpha_\lambda} \partial_t +
e^{-\beta_\lambda}\mathop{\sum_{i=1}^{d-2}}v^i\partial_{x_i}
\end{eqnarray}
with $\mathop{\sum_{i, j=1}^{d-2}} \tilde g_{ij}v^i v^j=1$ and
$v^i(x)$ independent of $t,z$:
\begin{eqnarray}
\label{wperp} 0 \le R_{ab} k^a_\perp k^b_\perp=(d-2)e^{\alpha_{\lambda}-\beta_{\lambda}}\frac{d}{dz}[{A_\lambda} ], \\
\label{wpara} 0 \le R_{ab} k^a_\parallel
k^b_\parallel=e^{-\alpha_{\lambda}-(d-2)\beta_{\lambda}}\frac{d}{dz}[{B_\lambda}
]+e^{-2\beta_{\lambda}} \tilde R \geq0,
\end{eqnarray}
where ${A_\lambda} (z)$ and ${B_\lambda} (z)$ are again defined by
(\ref{AB}) and $\tilde R(x) =  \sum_{i,j=1}^{d-2} \tilde R_{ij} v^i
v^j$ is defined in terms of the Ricci tensor $\tilde R_{ij}$ of the
metric $\tilde g_{ij}$.  Since (\ref{wperp}) is identical to
inequality (\ref{perp}) in section \ref{plane}, we see that, for
sufficiently small $\lambda$, the functions $e^{-\beta_{\lambda}}$
and $A_{\lambda}$ are bounded both above and below on the interval
$[z_-,z_+]$.

However, inequality (\ref{wpara}) differs from (\ref{para}) by the
addition of the final term.  Recall that our argument in section
\ref{plane} considered only the dependence on the coordinate $z$,
effectively taking the $x^i$ to be fixed, and note that $\tilde R$
is a function only of the $x^i$. Below, we again take the $x^i$ to
be fixed so that we may regard $\tilde R$ as a constant.
Furthermore, it is sufficient to consider the case $\tilde R > 0$
below, as for $\tilde R \le 0$ the constraint (\ref{wpara})
immediately reduces to (\ref{para}) and the rest of the argument
follows as in section \ref{plane}.

To show that $e^{\beta_\lambda (z)}$ is (uniformly) bounded under
our current hypotheses, recall that we have already established that
there is some positive constant $N$ such that $-A_\lambda<N$ for
sufficiently small $\lambda$. Thus, using the explicit expression
for $A_{\lambda}$, we have
\begin{equation}
\frac{d}{dz}[e^{\beta_\lambda}]<N e^{\alpha_\lambda}.
\end{equation}
Integrating this equation yields
\begin{equation}
e^{\beta_\lambda(z)}<e^{\beta_\lambda(z_-)}+
N\mathop{\int_{z_-}^{z}}e^{\alpha_\lambda}dz',
\end{equation}
and since the right hand side is uniformly bounded for sufficiently
small $\lambda$, so is $e^{\beta_\lambda}$. Since $\beta_\lambda$ is
bounded below, there in fact exists some positive constants $K_\pm$
such that $e^{\pm \beta_\lambda}<K_\pm$ for sufficiently small
$\lambda$.

Now, for each $x = \{x^i\}$, consider $z \in [-\epsilon, \epsilon]$
for any $\epsilon< |z_\pm|$, where we will later take $\epsilon$ to
be small. In order to provide a substitute for Corollary 1, we wish
to prove that $e^{\alpha_{\lambda}-\beta_{\lambda}}$ is bounded on
$[-\epsilon, \epsilon]$ uniformly in $\lambda$ (for small
$\lambda$). Integrating (\ref{wpara}) one finds
\begin{eqnarray}
\label{B2}
 {B_\lambda} (z)\geq
{B_\lambda} (-\epsilon)-\tilde R\mathop\int_{-\epsilon}^{z}
e^{\alpha_{\lambda}+(d-4)\beta_{\lambda}}dz'.
\end{eqnarray}
Since ${B_\lambda} (-\epsilon)$ converges in the limit $\lambda
\rightarrow 0$, for small $\lambda$ we have ${B_\lambda} (-\epsilon)
\ge -b_\epsilon$ where $b_\epsilon$ is some positive constant.
Considering the integral in (\ref{B2}), let us note that since
$e^{\alpha_{\lambda}-\beta_{\lambda}}$ is smooth, it attains some
maximal value ${\rm M_\lambda}$ on the interval $[-\epsilon,
\epsilon]$.  Thus, the integrand $e^{(d-3) \beta_\lambda}
e^{\alpha_\lambda-\beta_\lambda}$ is bounded (above) by
$K_+^{d-3}{\rm M_\lambda} > 0$. Multiplying the result by
$e^{-(d-1)\beta_{\lambda}}< K_-^{d-1}$  and using $z < \epsilon$ and
the definition of $B_\lambda$ thus yields
\begin{equation}
\label{dbound}
\frac{d}{dz}(e^{\alpha_{\lambda}(z)-\beta_{\lambda}(z)})
\geq-K_-^{d-1}b_{\epsilon}- 2\epsilon \tilde RK_-^{d-1}K_+^{d-3}
{\rm M_\lambda}.
\end{equation}

For the next step, it will be convenient to define $f_\lambda =
e^{\alpha_{\lambda}-\beta_{\lambda}}$, the function with which we
are currently concerned. Furthermore, we take $z_0$ to be the point
in $[-\epsilon, \epsilon]$ where $f_\lambda(z_0) = {\rm M_\lambda}$.
If $z_0< \epsilon,$ then by the mean value theorem the derivative
$\frac{d}{dz} f_\lambda$ attains the value
$\frac{f_\lambda(\epsilon)-{\rm M_\lambda}}{\epsilon-z_0}$ somewhere
between $z_0$ and $\epsilon$, so that this value must also satisfy
the bound (\ref{dbound}). But, since $z_0 > - \epsilon$ we have
\begin{equation}
\label{mean} \frac{f_{\lambda}(\epsilon) - {\rm
M_\lambda}}{2\epsilon} \geq
\frac{f_\lambda(\epsilon)-f_{\lambda}(z_0)}{\epsilon-z_0} \geq
-K_-^{d-1}b_{\epsilon}-  2\epsilon \tilde R K_-^{d-1}K_+^{d-3} {\rm
M_\lambda}.
\end{equation}

Choosing $\epsilon^2 < 1/4\tilde RK_-^{d-1}K_+^{d-3}$, a bit of
algebra yields
\begin{equation}
\label{uniform} {\rm M_{\lambda}}\leq
\frac{f_{\lambda}(\epsilon)+2\epsilon
K_-^{d-1}b_{\epsilon}}{1-\tilde RK_-^{d-1}K_+^{d-3} 4\epsilon^2}.
\end{equation}
Note that (\ref{uniform}) also holds for the remaining case where
$f_\lambda$ attains the value ${\rm M_\lambda}$ only at
$z=\epsilon$.

Finally, since $f_\lambda(\epsilon)$ converges as $\lambda
\rightarrow 0$, the right hand side is bounded uniformly in
$\lambda$ for small $\lambda$.  We see that ${\rm M_\lambda}$ and
thus $e^{\alpha_{\lambda}-\beta_{\lambda}}$ (on $[-\epsilon,
\epsilon]$) has a $\lambda$-independent upper bound for small
$\lambda$. In summary, we have shown:

\begin{quote}
$  Lemma \ 3$: Given $\alpha_\lambda, \beta_\lambda$ satisfying
(\ref{wperp}), (\ref{wpara}), and the stated convergence properties
with respect to $\lambda$, there is an $\epsilon>0$ such that, for
sufficiently small $\lambda$, the functions $e^{-\beta_{\lambda}}$
and $e^{\alpha_{\lambda}- \beta_\lambda}$ are bounded (uniformly in
$\lambda$) on the interval $[-\epsilon, \epsilon]$.
\end{quote}

It is now straightforward to prove continuity of the induced metric.
As above, we consider only the interval  $[-\epsilon, \epsilon]$ and
the case $\tilde R > 0$. By Lemmas  2 and 3, for sufficiently small
$\lambda$, the function $\beta_{\lambda}' = -{A_\lambda}
e^{\alpha_{\lambda}-\beta_{\lambda}}$ is bounded uniformly in
$\lambda$ (for small $\lambda$) so that $\beta_0$ is continuous. In
addition, we have
\begin{equation}
\frac{d}{dz}[{B_\lambda}
]\geq-e^{\alpha_{\lambda}+(d-4)\beta_{\lambda}}\tilde R
\geq-e^{\alpha_{\lambda}-\beta_{\lambda}} K_+^{d-3} \tilde R \ge -
K_+^{d-3} \tilde R {\rm M_\lambda}.
\end{equation}
for sufficiently small $\lambda$.  Thus for $z \in [-\epsilon,
\epsilon]$ we have
\begin{equation}
{B_\lambda} (\epsilon) + 2\epsilon K_+^{d-3} \tilde R {\rm
M_\lambda} \ge {B_\lambda} (z) \ge {B_\lambda} (-\epsilon) -
2\epsilon K_+^{d-3} \tilde R {\rm M_\lambda}.
\end{equation}
The convergence of ${B_\lambda} (\pm \epsilon)$ as $\lambda
\rightarrow 0$ and  Lemma 3 then guarantee that on $[-\epsilon,
\epsilon]$, the function ${B_\lambda} =
 e^{(d-1)\beta_{\lambda}} \frac{d}{dz}(e^{\alpha_{\lambda}-\beta_{\lambda}})
$ satisfies a bound which is  uniform in $\lambda$ for small
$\lambda$.  Multiplying by $e^{-(d-1)\beta_{\lambda}}$ and using
Lemma 3 once again, the same must be true of
$\frac{d}{dz}(e^{\alpha_{\lambda}-\beta_{\lambda}})$.  Thus
$e^{\alpha_{0}-\beta_{0}}$ is continuous.  Since we have already
established continuity of $e^{\beta_{0}}$, we in fact have:

\begin{quote}
$   Theorem \ 2$: Consider a one-parameter family of smooth metrics
of the form (\ref{warped}) satisfying conditions (\ref{smooth}) and
(\ref{ncc}) from section \ref{intro}.  If null geodesics along
surfaces of constant $x^i$ in the limiting spacetime $(M,g_0)$ reach
(and cross) the wall in finite affine parameter, then the limiting
metric $g_0$ defines continuous induced metrics on the hypersurfaces
of constant $z$.
\\
\end{quote}

\subsection{More degrees of freedom}

\label{diagonal}

Thus far, we have investigated only metric ans\"atze having two free
functions.  However, it is also straightforward to consider metrics
of the form:

\begin{equation}
\label{diag} ds_{\lambda}^2 =
-dt^2+dz^2+\mathop{\sum_{i=1}^{d-2}}e^{2\beta_{\lambda}^i(z)}{(dx^i)^2}.
\end{equation}
Note that we have required the existence of a timelike Killing
vector field with constant norm.  As usual, we take the range of $z$
to include $[z_-,z_+]$, and we take this coordinate to label the
hypersurfaces $\Sigma^z$. Since no redshift factor is allowed in
(\ref{diag}), such metrics are perhaps of less physical interest
than those considered earlier. Nonetheless, we present a short proof
for this case in the hopes that it will prove useful for future
investigations.

The techniques developed in section \ref{plane} apply in a
straightforward way.  It is sufficient to consider the null vectors
\begin{equation}
k_i=\partial_t + e^{-\beta^i_\lambda}
\partial_{x_i},
\end{equation}
for which the null convergence condition implies
\begin{eqnarray}
\label{idiag} 0 \le R_{ab} k_i^a k_i^b =
-{\beta_{\lambda}^i}''-{\beta_{\lambda}^i}'(\mathop{\sum_{k=1}^{d-2}}{\beta_{\lambda}^k})'.
\end{eqnarray}
Repeating the arguments of section \ref{plane}, one immediately
finds that  $\beta_{\lambda}^i$ has no local minima on $[z_-,z_+]$
and thus is bounded below uniformly in $\lambda$ for sufficiently
small $\lambda$.

Furthermore, introducing
$B_{\lambda}^i:=-{\beta_{\lambda}^i}'e^{\mathop{\sum_{k=1}^{d-2}}\beta_{\lambda}^k}$,
one finds that  (\ref{idiag}) simplifies to
$0\leq\frac{d}{dz}[B_{\lambda}^i]$. Thus, repeating our standard
argument, $B_{\lambda}^i$ is bounded uniformly in $\lambda$ for
small $\lambda$.  Together, these facts imply a uniform bound on
${\beta_{\lambda}^i}' = -B_{\lambda}^i
e^{-\mathop{\sum_{k=1}^{d-2}}\beta_{\lambda}^k}$ for small
$\lambda$, which in turn implies continuity of $\beta_{0}^i$ in the
limit. We thus verify our conjecture for the special case defined by
the ansatz (\ref{diag}).

\section{A result without symmetry}
\label{congruence}

In this section we derive a result which ensures continuity of the
induced metric across $\Sigma^0$ without specifying an ansatz for
$g_\lambda$ and without imposing symmetry. However, to achieve this
we introduce a number of assumptions concerning the behavior of null
geodesics in the spacetime $(M,g_0)$. Our result below (Theorem 3)
states roughly that it is impossible for the induced metric to be
discontinuous across a positive energy wall unless some other
pathology also occurs.

To this end, we begin with a definition which introduces a notion of
convergence for a family $C_\lambda$ of geodesic congruences
associated with the spacetimes $(M,g_\lambda)$.

\begin{quote} $Definition \ 1$:
Consider a family of spacetimes $(M,g_\lambda)$ satisfying condition
(i) and let $C_\lambda$ be a geodesic congruence in $(M,g_\lambda)$.
We then say that $C_\lambda$ converges to a collection of curves
$C_0$ when the following conditions hold.

\begin{enumerate}[a)]

\item  Note that each congruence $C_\lambda$ consists of a set of
geodesics $\{ \gamma_\lambda^x \}$ for $x$ in some appropriate label
space $X$.  We require that there be a choice of affine parameter
$s$ along each curve $\gamma_\lambda^x$ such that, for each $x$ in
the label set $X$, the affine parameter ranges over the interval
$[0,1]$ and the maps $\gamma_\lambda^x(s): [0,1] \rightarrow M$
converge pointwise in the limit $\lambda \rightarrow 0$ to some map
$\gamma_0^x(s): [0,1] \rightarrow M$.

\item  Away from the hypersurface $\Sigma^0$, we require that the
first and second derivatives of $\gamma^x_\lambda(s)$ with respect
to both $x$ and $s$ must converge to the corresponding derivatives
of $\gamma^x_0(s)$, which we also require to be well-defined. Thus,
away from $\Sigma^0$, we see that $\gamma^x_0$ is also a geodesic of
$(M,g_0)$ and that $s$ is again an affine parameter.
\end{enumerate}

We denote the collection of functions $\gamma_0^x: [0,1] \rightarrow
M$ by $C_0$.
\end{quote}

As defined above, $C_0$ is merely a collection of curves and need
not form a congruence.  For example, it is possible that all of the
curves in $C_0$ might coincide. However, consider the case where the
$C_\lambda$ are hypersurface orthogonal null congruences (without
caustics) and where the spacetimes $(M,g_\lambda)$ satisfy the null
convergence condition (\ref{ncc}).  Then by the Raychaudhuri
equation, the expansion $\theta$ is non-increasing along each
geodesic in the congruence (see e.g., \cite{Wald}). But the
expansion is determined by mixed second derivatives of the functions
$\gamma_\lambda^x(s)$ with respect to $x$ and $s$ together with the
metric $g_\lambda$ and its first derivative. As a result, whenever
$\theta_{0,x}$ is well-defined the expansion must converge in the
limit $\lambda \rightarrow 0$ to the associated expansion
$\theta_{0,x}$ of $\gamma_0^x(s)$. Thus, in any interval where it is
well-defined, the expansion of $\gamma_0^x(s)$ is again a
non-increasing function of $s$. Furthermore, since the area elements
carried by the congruences $C_\lambda$ converge to the area element
carried by $C_0$, the expansion $\theta_{0,x}$ can cease to be
well-defined only at $\Sigma^0$ or when $\theta_{0,x} \rightarrow -
\infty$ at some affine parameter $s_0$.

Suppose then that $\gamma_0^x$ is not in $\Sigma^0$ at $s=1$ and
that $C_0$ does form a congruence near $s=1$, so that the
expansion $\theta_{0,x}$ is well-defined there. Then, since
$\theta_{\lambda,x}|_{s=1}$ converges as $\lambda \rightarrow 0$
to $\theta_{0,x}|_{s=1},$ for sufficiently small $\lambda$, the
expansions $\theta_{\lambda,x}|_{s=1}$ are bounded uniformly in
$\lambda$. But since each $\theta_{\lambda,x}(s)$ is a
non-increasing function, this means that such
$\theta_{\lambda,x}(s)$ are bounded below uniformly in both
$\lambda$ and $s$. As a result, the limiting expansion
$\theta_{0,x}(s)$ associated with $\lambda =0$ is also bounded
below and does not diverge to $-\infty$.

Similarly, if $\gamma_0^x|_{s=0}$ is not in $\Sigma^0$ and if $C_0$
forms a congruence near $s=0$, then $\theta_{0,x}(s)$ is bounded
above. If both of these conditions hold, then the expansion
$\theta_{0,x}(s)$ is well-defined everywhere away from $\Sigma^0$
and is bounded.

It is useful to summarize this discussion in the following Lemma:

\begin{quote}$Lemma \ 4$:  Suppose that $C_\lambda$ are
hypersurface-orthogonal null congruences in the spacetimes
$(M,g_\lambda)$ which converge to the set of curves $C_0$ in
$(M,g_0)$ in the sense of Definition 1. Suppose also that
$\gamma_0^x|_{s=0},\gamma_0^x|_{s=1} \not \in \Sigma^0$ and that
$C_0$ forms a congruence near $s=0$ and near $s=1$. If the
spacetimes $(M,g_\lambda)$ satisfy the null convergence condition
(\ref{ncc}), then the expansion of $C_0$ is well-defined away from
$\Sigma^0$ and is bounded along each curve in $C_0$.
\end{quote}

Now, we wish to show that the induced metric must be continuous
across $\Sigma^0$ unless some additional pathology arises in the
spacetime.  Stating this precisely will also require us to define
a notion of the induced metric $h$ on {\it each} side of the
hypersurface $\Sigma^0$ in the spacetime $(M,g_0)$. We do so as
follows:

\begin{quote} $Definition \ 2$:
Consider a singular spacetime $(M,g_0)$ defined as a limit of smooth
spacetimes $(M,g_\lambda)$ satisfying condition (i). Furthermore,
consider any smooth foliation of the manifold $M$ near the singular
hypersurface $\Sigma^0$ which includes $\Sigma^0$ as a leaf.  Let us
label the leaves by a parameter $z$ taking $z=0$ at $\Sigma^0$, and
referring to the leaf at $z$ as $\Sigma^z$.

Now, for each $z > 0$ there is some metric $h_{z}$ induced by
embedding $\Sigma^z$ in $(M,g_0)$.  Suppose that the limit $h_{0+}
:=\lim_{z \downarrow 0} h_z $ (in which $z$ approaches zero from the
positive side) exists as a tensor field on $\Sigma^0$. Then we say
that $h_{0+}$ is an induced metric on the $z> 0$ side of $\Sigma^0$.
We may similarly define an induced metric $h_{0-}$ on the $z< 0$
side of $\Sigma^0$.
\end{quote}
Note that, in general, $h_{0+}$ need not agree with $h_{0-}$; i.e.,
the induced metric need not be continuous across $\Sigma^0$. Also
note that the definition above allows $h_{0\pm}$ to depend on the
choice of foliation.

We are now ready to state our theorem:

\begin{quote}
$  Theorem \ 3$: Consider any spacetime $(M,g_0)$ satisfying (i) and
(ii) from section \ref{intro} and a foliation $\{\Sigma^z \}$ of $M$
near $\Sigma^0$ which induces a well-defined, invertible, Lorentz
signature metric $h_{0\pm}$ on each side of $\Sigma^0$. Suppose also
that for each point $p \in \Sigma^0$ and each $\omega$ in some open
set of rank $(d-2)$ antisymmetric contravariant tensors associated
with the tangent space to $\Sigma^0$ at $p$, there is a set of
curves $C_0$ such that
\begin{enumerate}[a)]
\item $C_0$ satisfies the conditions of Lemma 4.

\item Some curve in $C_0$ passes through $p$.

\item  The tensor $\omega$ is the antisymmetric product of
the set of deviation vectors associated with $C_0$ at $p$; i.e, if
$y^a$ are coordinates on $\Sigma^0$, then
\begin{equation}
\label{omdef} \omega^{a_1,...,a_{d-2}} =
\epsilon^{i_1,...,i_{d-2}}y^{a_1}_{,i_1}...y^{a_{d-2}}_{,i_{d-2}},
\end{equation}
where the derivatives are taken with respect to coordinates $x^i$ on
the label space $X$ used to parameterize the curves in $C_0$. In
particular, we require the right-hand side of (\ref{omdef}) to be
well-defined; i.e., the analogue of (\ref{omdef}) for $\Sigma^z$
with $z \neq 0$ is continuous in $z$.

\item For each curve $\gamma_0^x$ in $C_0$ the point $\gamma_0^x(s)$ lies in $\Sigma^0$ for exactly one
value of the parameter $s$. This feature provides a sense in which
the wall may be considered ``thin.''
\end{enumerate}
Then the induced metric is in fact continuous across $\Sigma^0$;
i.e., $h_{0+} = h_{0-}$.
\end{quote}

The proof is straightforward. For any such $C_0$, consider the
infinitesimal area element carried by a curve $\gamma_0^x$ lying in
$C_0$ and running through $p$. This infinitesimal area element is
the antisymmetric product of the deviations vectors analogous to
(\ref{omdef}), but with derivatives taken holding fixed the
parameter $s$ along the curves. However, since the curves in $C_0$
are null away from $\Sigma^0$, the area of this element at any point
$q \not \in \Sigma^0$ may instead be evaluated using the deviation
vectors along a surface of constant $z$. That is, one may use
(\ref{omdef}) on a surface $\Sigma^z$ containing $q$ (and thus with
$z \neq 0$).

In particular, we may evaluate the limits of the area as $z
\rightarrow 0$ from either side. The result is simply the area
assigned to the corresponding rank $(d-2)$ antisymmetric tensor
$\omega$ at $p$ by the limiting metric $h_{0+}$ or $h_{0-}$.

Now, by assumption, the metrics $h_{0\pm}$ are non-degenerate.
Thus, knowledge of such areas for an open set (i.e., a
$(d-1)$-parameter family) of rank $(d-2)$ contravariant tensors
$\omega$ on $\Sigma^0$ in fact determines the full induced
metric\footnote{This follows from the fact that the square of the
area $A$ associated with the antisymmetric contravariant tensor
$\omega$ of rank $(d-2)$ is $A^2(\omega) = \frac{1}{(d-2)!}
\omega^{a_1...a_{d-2}} \omega^{b_1...b_{d-2}} h_{a_1 b_1} ...
h_{a_{d-2}b_{d-2}}.$  Thus, we have $\frac{1}{2}
\epsilon^{a_1...a_{d-2} a} \epsilon^{b_1...b_{d-2} b}
\frac{\partial^2 A^2}{\partial \omega^{a_1...a_{d-2}}
\partial \omega^{b_1...b_{d-2}}} = (det \ h) h^{ab},$ since $h_{ab}$ is invertible.}\textbf{}
on $\Sigma^0$. But these areas are continuous functions of the
affine parameter by Lemma 4, and since we have imposed that the
wall is ``thin'' as measured by $s$, this affine parameter is a
continuous function of $z$. Thus, the induced metrics on the
leaves $\Sigma^z$ must also be continuous in $z$. In particular,
we must have $h_{0+}=h_{0-}$.

\section{Discussion}
\label{disc}

In the work above, we have explored the idea that the null
convergence condition might imply continuity of the induced metric
across thin walls in general relativity and thus lead directly to
the Israel junction conditions \cite{Israel}. A full proof of this
conjecture would solidify the physical basis behind Israel's
formalism, whereas a counter-example would lead to a new class of
physically interesting thin-wall solutions.

It is important to stress, however, that in cases where one does not
require the null convergence condition we see {\it no} general
physical justification for requiring continuity of the induced
metric, or indeed any other simple behavior at the thin-wall
singularity.  Thin-wall limits of general smooth negative energy
spacetimes will thus be quite complicated to describe.  For example,
the Riemann tensor will in general fail to be a well-defined
distribution due to the non-linearities in its definition.  Thus, it
is not clear that the limiting thin-wall spacetime should itself
satisfy any well-defined version of the Einstein equation;
certainly, it need not satisfy the Israel conditions \cite{Israel}.
As a result, there may well be a large new class of thin-wall
spacetimes not satisfying the Israel conditions \cite{Israel} but
which are of physical relevance to the Randall-Sundrum scenario
\cite{RS1,RS2}, as this scenario violates the null convergence
condition.

Returning now to the present work, we have proven that the null
convergence condition leads to continuity of the induced metric for
various special cases. In section \ref{proof}, we considered thin
walls which can be approximated by certain families of smooth warped
product spacetimes; in particular, we considered approximating
metrics $g_\lambda$ of the form

\begin{equation}
ds^2 = -e^{2\alpha (z)}dt^2+dz^2+e^{2\beta(z)}\mathop{\sum_{i,
j=1}^{d-2}}\tilde g_{ij}(x){dx^i}{dx^j}.
\end{equation} The physics
of this ansatz can be decomposed into a number of independent
assumptions.  First, we have taken the approximating spacetimes to
be independent of time.  While this simplifies the calculations, one
expects a similar proof to go through in the general time-dependent
case.  The point is that, in the limit of a {\it smooth} thin wall,
it is natural to require derivatives along the wall (e.g., time
derivatives) to remain small compared to derivatives across the wall
(which necessarily become large). Thus, allowing non-zero tangential
derivatives should be only a small correction to the equations of
motion when viewed relative to the large transverse derivatives.

However, other generalizations (such as allowing off-diagonal
components of $g_\lambda$) may be more subtle, and seem unlikely to
be tractable using only the rather elementary methods of section
\ref{proof}.  One might hope to gain more control by following
Israel \cite{Israel} and casting the problem in terms of the metric
and intrinsic curvature of some foliation. However, our attempts to
date in this direction have not proven fruitful. We therefore leave
a more complete analysis for further investigations.

It is interesting to contrast the use of metric ans\"atze (Theorems
1 and 2 in section \ref{proof}) with the approach based on null
congruences
 used to prove Theorem 3 in section \ref{congruence}. On the one
hand, Theorem 3 is quite powerful as it requires neither symmetries
nor any other particular ansatz form for the approximating
spacetimes $(M,g_\lambda)$.  On the other hand, the reader will note
that the complete list of assumptions required to prove Theorem 3 is
rather long.  In particular, recall that Theorem 3 assumes not only
that the induced metric on each side of the wall exists as a tensor
field, but also that it is invertible and of Lorentz signature.
This is in contrast to the results of section \ref{proof} which
prove (within their own context) that the metric on each side of the
wall is well-defined.

Furthermore, the settings in section \ref{proof} include cases in
which the induced metric on the thin wall is degenerate; i.e., where
the wall becomes null in the limit $\lambda \rightarrow 0$. In
section \ref{proof} we were able to show only that the function
$e^{\alpha_0}$ is continuous at $\Sigma^0$, allowing $\alpha_0$
itself to diverge toward negative infinity as $z \rightarrow 0$, in
which case the induced metric at $\Sigma^0$ would be degenerate.  In
contrast, we were able to show that the function $\beta_0$ is
continuous at $\Sigma^0$, so that the corresponding metric component
$e^{\beta_0}$ is both continuous and non-vanishing.    Null walls
are quite physical, and a particularly interesting example arises in
thin-wall models of flat fundamental domain walls, whose stress
tensor is Poincar\'e invariant along the wall.  For example, we may
note from solutions \cite{DW2,DW3,DW4,DW1} for spherical such domain
walls that the walls tend to become null in the flat limit; i.e., as
the radius of the sphere grows to infinite size.  On the other hand,
the degenerate metrics associated with null walls cannot be
reconstructed from their area elements as would be required by a
generalization of Theorem 3.

Another assumption used in Theorem 3 was that one can identify an
open set (in particular, a $(d-1)$-parameter family) of null
geodesic congruences in each $(M,g_\lambda)$ such that, for each
value of the $(d-1)$ parameters, the associated congruences
$C_\lambda$ converge to a sufficiently nice family of curves $C_0$
in $(M,g_0)$. While this need not be the case for the situations
considered in section \ref{proof}, we note that all cases studied in
section \ref{proof} do have at least {\it one} such family of
congruences $C_\lambda$ originating on each side of the wall;
namely, the congruence of null geodesics with $x^i= constant$. It
would be interesting to understand if this is related to the
difficulty in generalizing the results of section \ref{proof} to
cases with non-trivial redshift and more than two free functions.

\begin{acknowledgments}  The authors thank Belkis Cabrera Palmer, Andr\'es Gomberoff, Eric Poisson, Ricardo Troncoso, Jorge Zanelli, and  especially Werner Israel for useful discussions.
The original idea for this project arose in discussions between one
of the authors (DM) and Belkis Cabrera Palmer. This work was
supported in part by NSF grant PHY0354978 and by funds from the
University of California and the Perimeter Institute for Theoretical
Physics. Finally, one of the authors (SY) would like to express his
gratitude to the UCSB College of Creative Studies for providing a
flexible educational system, from which he benefited enormously.
\end{acknowledgments}

\end{document}